\documentclass[showpacs,preprintnumbers,%
amsmath,amssymb,prc]{revtex4}
\usepackage{graphicx}
\usepackage{bm}
\begin{document}

\title{Final state polarization of protons in $pp \to pp \omega$}
\author{G. Ramachandran}
 \affiliation{GVK Academy, Jayanagar,Bangalore-560082, India}
\author{Venkataraya}
\altaffiliation{GVK Academy, Jayanagar,Bangalore-560082, India}
 \affiliation{Vijaya College, Bangalore-560011, India}
\author{J. Balasubramayam} 
   \altaffiliation{GVK Academy, Jayanagar,Bangalore-560082, India}
  \affiliation{K.S. Institute of Technology,  Bangalore, 560062, India}

\date{\today}
\begin{abstract}
Model independent formulae are derived for the polarizations and  spin correlations of protons
in the final state of $pp \to pp\omega $, taking 
into consideration all the six threshold partial wave amplitudes $f_1, \ldots , f_6$ covering 
$Ss, Sp$ and $Ps$ channels.   It is shown that these measurements of the final state spin observables, 
employing  only an unpolarized beam and an unpolarized target, may be utilized to complement 
measurements, at the double differential level, suggested earlier [Phys. Rev. \textbf{C78}, 01210(R)(2009)]
so that all the six partial wave amplitudes may be determined empirically.  
\end{abstract}


\maketitle

Meson production in $NN$ collisions has attracted considerable attention \cite{mach}  since the early 1990's, when 
total cross-section measurements \cite{meyer} 
for neutral pion production were found to be  more than a factor of 5 than the then available theoretical 
predictions \cite{koltun}. Experimental studies have indeed reached a high degree of sophistication since then and detailed 
measurements of the differential cross-section and  of spin observables have been carried out  employing  a polarized
beam and a polarized target \cite{preze,meyer63}. Apart from the pseudoscalar pion, vector mesons are also known to be 
significant contributors for the $NN$ interaction. When a meson is produced in the final state, a large momentum 
transfer is involved, which implies that the  $NN$ interaction is probed at very short distances,
 estimated \cite{nakayama}  to be of the order of 
$0.53fm$, $0.21fm$ and $0.18fm$ for the production of  $\pi, \omega$,  
and $\varphi$  respectively. Since, the singlet-octet mixing angle is close to the ideal value, the $\omega$  meson wave 
function is dominated by $u$ and $d$ quarks while the strange quark dominates in the case of  $\varphi$.
As a result, the  $\varphi$  meson production is suppressed as compared to the $\omega$   meson production, 
according to the Okubo-Zweig-Iizuka (OZI) rule \cite{okubo} . This rule was, however, found to be violated 
dramatically in the case of  $p \bar{p}$ collisions \cite{amsler} . Consequently,  attention has been focused 
on the measurement  \cite{balestra1,balestra2,hartmann}  of the ratio $R_{\varphi/\omega}$  and it's comparison with the 
theoretical estimates \cite{lipkin}. Apart from \cite{balestra1,balestra2,hartmann}  
measurements of total cross-section as well as angular distributions for $pp \to pp \omega$  
\cite{hibou}  at energies $\epsilon$  above threshold up to $320 MeV$ in c.m., the reaction has 
also been studied theoretically using several models \cite{koniuk} .  A model independent theoretical approach has also been 
developed  \cite{gr1}  to study the measurements of not only  the differential and total cross-sections, 
but also the polarization of  $\omega$ in the final state. A set of six partial wave amplitudes 
$f_1, \ldots, f_6$ have been identified \cite{gr1} to study $pp \to pp \omega$  at threshold and near threshold 
energies covering the  $Ss, Sp$ and $Ps$ amplitudes.  It was further shown   \cite{gr2} that the dominant decay mode   
$\omega \to 3\pi$  can only be utilized   to determine the tensor polarization of $\omega$. On the 
other hand, it was also pointed out \cite{gr1} that  the vector as well as tensor poalrizations can be measured using
the decay $\omega \to \pi^0 \gamma$,  with the smaller branching ratio of $8.92\%$. 
It is encouraging to note that WASA \cite{was}  at COSY  is expected to facilitate the experimental study of 
$pp \to pp\omega$ via the detection of $\omega \to \pi^0 \gamma$ decay.
In view of a recent measurement \cite{abd}  of the analyzing power $A_y$ for the first time, 
the model independent approach was extended to \cite{jb}   
study $\omega$   production in $pp$-collisions with a polarized beam.
While considering $\omega$ production it is worth pointing out that the notation used by 
Meyer et al., \cite{meyer63} in the context of neutral pion production 
has to be complemented. Since $\omega$
is a spin 1 meson, one needs to specify also the total angular momentum 
$j_\omega = |l-1|,\ldots ,l+1$  of the $\omega$ meson where $l$ denote the orbital angular 
momentum with which the meson is produced. Moreover $j_\omega$ has to combine with $j_f$ of the two nucleon system 
in the final state to yield total angular momentum $j$ of the two 
nucleon system in the initial state due to the rotational invarience. This problem has been discussed in
\cite{jb} and the amplitudes $f_1, \ldots , f_6$ have explicitly been given in terms of the amplitudes
which specify $j_f$ and $j_\omega$. 
Considering the beam analyzing power $A_y$ and beam to meson spin transfers in addition to the
differential cross-section, at the double differential level,
 it was shown in \cite{jb} that the lowest three amplitudes $f_1,f_2,f_3$
covering the $Ss$ and $Sp$ channel can be determined empirically without any discrete ambiguity, while 
information with regard to the amplitudes $f_4,f_5,f_6$  covering the $Ps$ channel can only be extracted 
partially from these measurements.

The purpose of the   present paper is to demonstrate theoretically that all 
the six amplitudes may be determined empirically without any ambiguities, 
if some measurements are carried out  with regard to the final 
spin state of the protons  in an experiment employing an unpolarized beam and an unpolarized target.
We may perhaps mention here that we do not make any simplifying assumptions as has been made in an older work \cite{gr3}.
It may further be noted that the analysis reported in \cite{gr3} made use of the then existing unpolarized 
differential cross-section measurements at the single differential level, whereas we are considerng here 
all the observables at the double differential level as in our more recent work \cite{jb}. As such the present paper carries forward 
the analysis reported in \cite{jb} and is not in any way dependent on the much earlier results of \cite{gr3}.

The reaction matrix $\mathcal{M}$ may  be expressed, in a model independent way \cite{gr1,gr3,jb}, as
\begin{eqnarray}
\label{tm}
\mathcal{M} & = &
\sum_{s_f,s_i=0}^1 \sum_{\lambda=|s_f-s_i|}^{(s_f+s_i)}
\sum_{S=1-s_f}^{1+s_f}\sum_{\Lambda = |S-s_i|}^{(S+s_i)} \nonumber \\ 
& & \times(( S^1(1,0) \otimes S^{\lambda}(s_f,s_i))^{\Lambda} \cdot 
\mathcal{M}^{\Lambda}(Ss_fs_i;\lambda)),
\end{eqnarray}
where $s_i,s_f$ denote respectively the initial and final spin states of the two nucleon system and $S$, the 
channel spin in the final state of the reaction. The irreducible tensor amplitudes $\mathcal{M}^{\Lambda}_{\nu}
(Ss_fs_i;\lambda)$ of rank $\Lambda$ are explicitly given, in terms of partial wave amplitudes
$f_1, \ldots , f_6$,  by 
\begin{align}
{\cal  M}_{0}^{1}(101;1) &= \frac{1}{24 \pi \sqrt{\pi}} f_1, \\
{\cal  M}_{\pm 1}^{1}(101;1) &= 0, \\
{\cal  M}_{0}^{1}(100;0) &= \frac{1}{8\pi \sqrt{3\pi}} 
f_{23}'\,cos\theta,\\
{\cal  M}_{\pm 1}^{1}(100;0) &= \mp \frac{1}{8\pi \sqrt{6\pi}} 
f_{23} \,sin\theta e^{\pm i \varphi}, \\
{\cal  M}_{0}^{1}(110;1) &= \frac{1}{8\pi \sqrt{3\pi}} f_{45}'
\,cos\theta_f,\\
{\cal  M}_{\pm}^{1}(110;1) &= \mp \frac{1}{8\pi \sqrt{6\pi}} f_{45} \,sin\theta_f, \\
{\cal  M}_{0}^{2}(210;1) &= 0,\\
{\cal  M}_{\pm 1}^{2}(210;1) &= - \frac{3}{80\pi \sqrt{3 \pi} } f_6 \,sin\theta_f ,\\
{\cal  M}_{\pm 2}^{2}(210;1) &= 0,
\end{align}
where the $z$-axis is chosen along the beam, and the plane containing the beam and $ \boldsymbol{p_f = (p_1-p_2)}/2 $
is chosen as the $z$-$x$ plane if $\boldsymbol{p_1}$ and $\boldsymbol{p_2}$ denote c.m. momenta of the two protons in the final state.
The polar angles of the c.m. momentum of the meson are denoted  by $(\theta, \varphi)$. The shorthand notation 
\begin{eqnarray}
f_{ij}  &=&  f_i + \frac{1}{\sqrt{10}}f_j, \\
f_{ij}'  &=&  f_i - \frac{2}{\sqrt{10}}f_j,
\end{eqnarray}
is used with $i,j = 2,3 \; \mbox{ or} \; 4,5$.

When the beam and target are unpolarized the spin density matrix $\rho^f$ characterizing the final spin 
state of the system is given by 
\begin{equation}
\rho^f = \frac{1}{4}{\cal M}{\cal M}^\dagger,
\end{equation}
so that the unpolarized double differential cross-section is given by 
\begin{eqnarray}
\label{updcs1}
\frac{d^2\sigma_o}{dW d\Omega_f d\Omega} &=&   
 Tr(\rho^f)   \equiv d^2\sigma_0,
\end{eqnarray}                                                                                    
where the abbreviation $d^2\sigma_0$ is employed for simplicity.

If measurements are not carried out with 
regard to the spin state of $\omega$, the density matrix 
\begin{equation}
\rho = \sum_{\mu=-1}^{1}\langle 1 \mu |\rho^f |1 \mu \rangle ,
\end{equation}
describes the spin state of the protons in the final state. Here $|1 \mu \rangle$ denotes
the spin state of $\omega$, with magnetic quantum number $\mu$.

It is well known that the state of polarization  of  two protons 
is completely  specified by measuring the expectation values 
\begin{eqnarray}	
\label{palpha}   
d^2\sigma_0\,P_\alpha(i) &=& Tr[\sigma_\alpha(i) \rho ],\,\,\, i = 1,2,
\end{eqnarray}
which denote the individual polarizations of the two protons in the final state 
and  their spin correlations  
\begin{eqnarray}	
\label{calpha} 
d^2\sigma_0\,C_{\alpha, \beta} &=& Tr[\sigma_\alpha(1)\sigma_\beta(2)\rho],
\end{eqnarray}
where $\alpha, \beta$ denote Cartesian components $\alpha, \beta = x, y , z$.                        
We obtain
\begin{eqnarray}
\label{px1}
 -P_x(1) = P_x(2) &=&  \;g \;\Im(\gamma)  \nonumber \\  & \times &   sin\theta
sin\varphi cos\theta_f ,\\ 
\label{py1}                                              
P_y(1) &=&  \; g \; [ \frac{\sqrt{3}}{2\sqrt{2}} \Im(\eta_3) \nonumber \\ && -   
\Im(\gamma)sin\theta cos\varphi cos\theta_f \nonumber \\ && + 
\Im(\eta_2)cos\theta  sin\theta_f ], \\ 
\label{py2}  
P_y(2) &=&  \; g \; [ \frac{\sqrt{3}}{2\sqrt{2}} \Im(\eta_3) \nonumber \\ && +   
\Im(\gamma)sin\theta cos\varphi cos\theta_f \nonumber \\ && -
\Im(\eta_2)cos\theta  sin\theta_f ], \\ 
\label{pz1}                                                 
-P_z(1) = P_z(2)  &=&   \;g \;\Im(\eta_1)  \nonumber \\  & \times &   sin\theta
sin\varphi sin\theta_f,  \\  
\label{cxy}                            
C_{xy} - C_{yx} &=&  2  \;g \;\Re(\eta_1)  \nonumber \\  & \times &   sin\theta
sin\varphi sin\theta_f , \\
\label{cyz}
C_{yz} - C_{zy} &=& -2  \;g \;\Re(\gamma)  \nonumber \\  & \times &   sin\theta
sin\varphi cos\theta_f ,\\ 
\label{czx}
C_{zx} - C_{xz} &=& 2  \;g \; \nonumber \\  & \times & [\Re(\gamma) sin\theta
cos\varphi cos\theta_f  \nonumber \\  & - & \Re(\eta_2) cos\theta
 sin\theta_f],
\end{eqnarray}              
where 

\begin{eqnarray}
\label{n0}
\gamma  &=& f_{23} f_{45}^{'*},\\
\label{n1}
\eta_1 &=& f_{23}( f_{45} - \frac{3}{\sqrt{50}} f_6 )^* , \\                                                                  
\label{n2}
\eta_2 &=& f_{23}'( f_{45} + \frac{3}{\sqrt{50}} f_6)^* , \\ 
\label{n3}                                                               
\eta_3 &=& f_{45}'( f_{45} - \frac{3}{\sqrt{50}} f_6)^*.
\end{eqnarray} 
and 
\begin{equation}
g = \frac{\sqrt{6}/32\pi^3} { Tr(\rho^f)}
\end{equation} 
is known from Eq.\eqref{updcs1}. The above formulae \eqref{px1} to \eqref{czx} for all the proton spin observables 
in the final state are derived for the first time. These observables at the double differential 
level complement the observables at the double differential  considered in \cite{jb}.

Experimental measurements of \eqref{cyz} and \eqref{px1} determine respectively real and imaginary parts 
of $\gamma$ given by \eqref{n0}. Likewise, \eqref{cxy} and \eqref{pz1} determine respectively the real and 
imaginary parts of $\eta_1$. Since $\Re({\gamma})$ is known from \eqref{cyz}, the real part of $\eta_2$ 
may be determined  from \eqref{czx}. If we consider $P_y(1)-P_y(2)$, it is clear on using \eqref{py1} and \eqref{py2}
that  $\Im(\eta_2)$ can be
determined, since $\Im(\gamma)$ is known from \eqref{px1}.
Taking into consideration these additional inputs together with inputs derived from the measurements
discussed earlier in \cite{jb}, it is possible to determine all the six partial wave amplitudes $f_1, \ldots , 
f_6$ along with their relative phases empirically.

\begin{table}[b]
	\centering
	\caption{Observables in $pp \to pp\omega$ at double differential level}
		\begin{tabular}{clc}\hline \hline
		    Sl. & Observables and their  &      Entities determinable  \\ 
		  No. & theoretical fomulae   & from  experimental   \\
		  &    &  measurements  \\ \hline
		 1& Unpolarized double differential  &  \\
		  & cross-section ;\;    $d^2\sigma_0=$               & $ a = (\alpha_0 + 9\zeta_0),  $ \\
		  & $\frac{1}{768\pi^3}[a+0.9\alpha_2cos^2\theta+9\zeta_2 cos^2\theta_f]$& $\alpha_2, \zeta_2$\\
		 2 & Vector polarization  & \\
		 & of $\omega ;\;  \;\;\; C_0(t^1_{\pm 1})_0 = $ & $\alpha_3, \;\; \zeta_3$ \\
		 & $\frac{9i}{4}[\frac{2}{\sqrt{10}}\alpha_3 sin2\theta + \zeta_3 sin2\theta_f e^{\pm i\varphi_f}]$&\\
		 3 & Tensor polarization of $\omega$ ;\;  &     \\
		  & $C_0(t^2_{0})_0 = $ &  $ b = (\alpha_4 - 9 \zeta_4)$ , \\ 
		  & $\frac{1}{\sqrt{6}}[b - 9 \alpha_5cos^2\theta+\zeta_5 cos^2\theta_f]$ & $ \alpha_5, \;\;\; \zeta_5 $ \\
		  & $ C_0(t^2_{\pm 1})_0 =$ &  $\alpha_6, \;\;\; \zeta_6$ \\& $\pm \frac{3}{4}[2\alpha_6 sin2\theta - 3 \zeta_6 sin2\theta_f
		   e^{\pm i\varphi_f}]$& \\
		  &  $C_0(t^2_{\pm 2})_0 = $ & $\alpha_7, \;\;\; \zeta_7 $ \\ & $ - \frac{3}{4}[2\alpha_7 sin^2\theta -
		  3 \zeta_7 sin^2\theta_f e^{\pm 2i\varphi_f}]$ & \\
		  4 & Beam analyzing power ;\;  & \\
		    & $C_0 \vec{A} = \sqrt{2} \beta_1 (\hat{q} \times \hat{p}_i)$ & $\beta_1$ \\
		  5 & Beam to $\omega$ spin transfers ;\;  & \\
		    & $C_0 K^x_x = C_0 K^y_y = -\beta_4 cos\theta,$ & $\beta_4$ , \\
			& $ C_0 K^z_x = \sqrt{2} \beta_2 sin\theta $ & $\beta_2$,  \\
			& $ C_0 K^z_z = \frac{1}{\sqrt{3}} \beta_3 $ & $\beta_3 , $ \\
			& $C_0 K^{xx}_y = -2 C_0 K^{yy}_y = -2 C_0 K^{zz}_y  $ & \\
			& $ = -2\sqrt{2}\beta_1 sin\theta $ & $ \beta_1$ \\
			& $ C_0K^{xz}_y = -C_0 k^{yz}_x = -\frac{3}{\sqrt{2}}\beta_5 cos\theta$ & $\beta_5 $\\
		  6 & Final state polarization of & $ \eta_1 , \eta_2 $ \\
		    & two protons  Eqs. $(18)$ to $(23)$   & $\eta_3 $ and $\gamma$ \\
			& of the present paper & \\
		  \hline  
		\end{tabular}
	\label{fkrk}
\end{table}

Let us therefore summarize in Table.\ref{fkrk} the information obtainable from various observables at the double 
differential level. We consider the unpolarized differential cross-section, polarization of $\omega$ produced,
the beam analysing power, the beam to $\omega$ meson spin transfers and the final state spin observables of 
the $pp-$system, formulae for which have been derived for the first time in this paper. 
The $\alpha, \beta, \zeta, \eta $ and $\gamma$ are bilinears in $f_1, f_{23}, f_{23}' , f_{45}, f_{45}' $ and
$f_6$. The explicit forms for $\eta_1, \eta_2, \eta_3$ are given by Eqs. \eqref{n1} to \eqref{n3}, while 
$\gamma$ is given by Eq.\eqref{n0} of the present paper. The explicit form for $\alpha_0 = \alpha_4$ is 
given by Eq.(7) and Eq.(19) of \cite{jb}.  We may re-write  $\alpha_2, \alpha_3, \alpha_5$
and $\alpha_6$ given by Eqs. (7), (19), (20)  and (21) of  \cite{jb} as 
\begin{eqnarray}
\label{al2}
\alpha_2 &=& |f_3|^2-2\sqrt{10}\Re(f_2f_3^*)=\frac{10}{3}(|f_{23}'|^2-|f_{23}|^2)\\
\label{al3}
\alpha_3 &=& \Im(f_2f_3^*)= -\frac{\sqrt{10}}{3}\Im(f_{23}f_{23}^*)\\
\label{al5}
\alpha_5 &=& |f_2|^2+ \frac{3}{10}|f_3|^2-\frac{2}{\sqrt{10}}\Re(f_2f_3^*) \nonumber \\  & = & \frac{1}{3}(|f_{23}|^2+2|f_{23}'|^2)\\
\label{al6}
\alpha_6 &=& |f_2|^2- \frac{1}{5}|f_3|^2-\frac{1}{\sqrt{10}}\Re(f_2f_3^*) = \Re(f_{23}f_{23}^*)
\end{eqnarray}

The explicit forms for  $\beta_1, \ldots , \beta_5$ are given in Eqs. (12), (37) and (38) of  [19] , while  
those for $\zeta_0, \zeta_2, \ldots , \zeta_7$ are given by Eqs. (8), (22),...,(26) of  [19]. 

We readily find that
\begin{equation}
\label{beta3}
|f_1|^2 = \beta_3
\end{equation}

   We may choose the phase of $f_1$ to be zero without any loss of generality so that $f_1$ is known empirically 
   from Eq. \eqref{beta3}. We denote the relative phases of  $f_{23}, f_{23}', f_{45}, f_{45}'$ and $f_6$
   with respect to $f_1$ as  $\varphi_{23}, \varphi_{23}', \varphi_{45}, \varphi_{45}'$ and $\varphi_6$ respectively. 
   We readily see that
 \begin{equation}
\label{f23}
|f_{23}|^2 = \alpha_7,
\end{equation}
whereas $\varphi_{23}$ is given, without any trigonometric ambiguity, by
\begin{equation}
\label{cos23} 
cos\varphi_{23} = \frac{\beta_2}{f_1 |f_{23}|} ;\;\hspace{1cm}\; sin\varphi_{23} = \frac{\beta_1}{f_1 |f_{23}|}
\end{equation}

Thus $f_{23}$ is known empirically. Likewise we find that
 \begin{eqnarray}
\label{f232}
|f_{23}'|^2 &=& \alpha_7+ 0.3 \alpha_2,\\
\label{cos232}
cos\varphi_{23}' &=& \frac{\beta_4}{f_1 |f_{23}'|} ; \;\hspace{1cm}\; sin\varphi_{23}' = -\frac{\beta_5}{f_1 |f_{23}'|}
\end{eqnarray}
which determine $f_{23}'$ empirically. Similarly
 \begin{eqnarray}
 \label{f45}
 |f_{45}|^2 &=& \left|\frac{f_{23}\eta_2+f_{23}'\eta_1}{2f_{23}f_{23}'}\right|^2, \\
 \label{cos45}
 cos\varphi_{45} &=& \frac{1}{2 f_1|f_{45}|} \nonumber \\ & \times & \left[ \frac{\beta_2\Re \eta_1 + \beta_1 \Im \eta_1}{|f_{23}|^2}
+ \frac{\beta_4\Re \eta_2 - \beta_5 \Im \eta_2}{|f_{23}'|^2} \right], \\
\label{sin45}
 sin\varphi_{45} &=& \frac{1}{2 f_1|f_{45}|}\nonumber \\ & \times & \left[ \frac{\beta_1\Re \eta_1 - \beta_2 \Im \eta_1}{|f_{23}|^2}
- \frac{\beta_5\Re \eta_2 + \beta_4 \Im \eta_2}{|f_{23}'|^2} \right],
\end{eqnarray}
determine $f_{45}$ empirically. We next note that 
\begin{equation}
\label{f45d}
|f_{45}'|^2 = \zeta_0 + \zeta_2 = \zeta_5 - \zeta_4,
\end{equation}
where 
\begin{equation}
\zeta_0 = \frac{1}{2}\zeta_7 + \frac{1}{27}(a-b)\;\;;\;\; \zeta_4 = -\frac{1}{2}\zeta_7 + \frac{2}{27}(a-b),
\end{equation}
in terms of the entities listed in the second column of Table.\ref{fkrk}. Moreover, 
\begin{equation}
\label{cos452}
 cos\varphi_{45}' = \frac{\beta_2\Re \gamma + \beta_1 \Im \gamma}{f_1|f_{45}'||f_{23}|^2}
 \;;\; sin\varphi_{45}' =   \frac{\beta_1\Re \gamma -\beta_2 \Im \gamma}{f_1|f_{45}'||f_{23}|^2}
 \end{equation}
 which together with \eqref{f45d} determine $f_{45}'$ empirically. Finally
 \begin{eqnarray}
 \label{f6}
 |f_{6}|^2 &=& \frac{25}{18}\left|\frac{f_{23}\eta_2-f_{23}'\eta_1}{f_{23}f_{23}'}\right|^2,\\
 \label{cos6}
 cos\varphi_{6} &=& \frac{5\sqrt{2}}{6 f_1|f_{6}|} \nonumber \\ & \times &\left[ \frac{\beta_4\Re \eta_2 - \beta_5 \Im \eta_2}{|f_{23}'|^2}
 - \frac{\beta_2\Re \eta_1 + \beta_1 \Im \eta_1}{|f_{23}|^2} \right], \\
 \label{sin6}
 sin\varphi_{6} &=& -\frac{5\sqrt{2}}{6 f_1|f_{6}|} \nonumber \\ & \times & \left[ \frac{\beta_5\Re \eta_2 + \beta_4 \Im \eta_2}{|f_{23}'|^2}
+ \frac{\beta_1\Re \eta_1 - \beta_2 \Im \eta_1}{|f_{23}|^2} \right],
\end{eqnarray}
which determine $f_6$ empirically. Thus we see from Eqs.\eqref{beta3}, \eqref{f23}, \eqref{f232}, \eqref{f45}
\eqref{f45d} and \eqref{f6} that the moduli of $f_1 , f_{23}, f_{23}', f_{45}, f_{45}'$ and $f_6$ can be 
determined. The relative phases of $f_{23}, f_{23}' f_{45}, f_{45}'$ and $f_6$ are determinable with respect 
$f_1$ using Eqs. \eqref{cos23}, \eqref{cos232}, \eqref{cos45}, \eqref{sin45}, \eqref{cos452}, \eqref{cos6}
and \eqref{sin6} without any trigonometric ambiguity, choosing $f_1$ to be real without any loss of genarality.
Therefore the amplitudes $f_1 , f_{23}, f_{23}', f_{45}, f_{45}'$ and $f_6$ are determinable purely empirically.

 It may be noted that $|f_1|$  is determined directly from a measurement of  the beam to meson spin 
   transfer $K_z^z$. The   $|f_{23}|$   and  $|f_{23}'|$   are determinable from the measurements of the unpolarized 
   differential cross-section and the tensor polarization of  $\omega$ The determination of relative phases of  
    $f_{23}$ and $f_{23}'$ with respect to $f_1$ involve measurement of beam to meson spin transfers. 
	The   $|f_{45}'|$ is determinable from unpolarized 
	differential cross-section and tensor polarization of $\omega$. The determination of relative phases   
	$\varphi_{45}'$ as well as $\varphi_6$ involve proton spin measurements in the final state which are 
	advocated for the first time in the present paper. 

Having determined $f_{ij}$ and $f_{ij}', i,j = 2,3$ or $4,5$ we may readily obtain $f_i$ and $f_j$ individually 
through
\begin{equation}
 \left( \begin{array}{c}
              f_i \\
			  f_j
              \end{array} \right) = \frac{1}{3}
\left( \begin{array}{cc}
              2 & 1 \\
			  \sqrt{10} & -\sqrt{10}
              \end{array} \right)
\left( \begin{array}{cc}
              f_{ij} \\
			 f_{ij}'
              \end{array} \right)		  
\end{equation}
 
   Thus, one can determine all the six partial wave amplitudes $f_1,\ldots, f_6$   purely empirically in terms of 
   entities ( listed in column 2 of Table.\ref{fkrk} ) which are extracted from the experimental measurements 
( listed in column 1 of Table.\ref{fkrk} ) at the double differential level.

\noindent
{\large{\bf Acknowledgments}}\\

One of us, (JB)  thanks Principal T. G. S. Moorthy and the Management of
K. S. Institute of technology  for encouragement and another (Venkataraya) 
acknowledges much encouragement for research given by the Principal  and 
the Management of Vijaya College.

\end{document}